\documentclass[11.5pt]{IEEEtran}
\pdfoutput=1
\usepackage{latexsym}
\usepackage{mathptmx}
\usepackage{amsfonts}
\usepackage{amssymb}
\usepackage{amsbsy}
\usepackage{amsthm}
\usepackage{textcomp} 
\usepackage{graphicx} 
\usepackage{amsmath} 
\usepackage{algorithm2e} 
\usepackage{tabu} 
\usepackage{caption}
\usepackage{subcaption}
\usepackage{wrapfig}
\usepackage{blindtext}
\usepackage{multicol}

\usepackage[pdftex,colorlinks=true,urlcolor=blue,citecolor=black,anchorcolor=black,linkcolor=black]{hyperref}

\begin{document}
\title{Challenges of Growing Social Media Networks From the Bottom-Up Through the Agent Perspective}
\author{Joseph A.E. Shaheen \\ [12pt]
Department of Computational and Data Science \\
George Mason University \\
4400 University Drive, MS 6A2 \\
Fairfax, VA 22030-4444, USA \\
jshaheen@gmu.edu \\}

\maketitle

\section*{Abstract}

We develop an agent-based model in order to understand agent/node behaviors that generate social media networks. We use simple rules to synthetically generate a backcloth (friend/follow) network collected using Twitter's API. The Twitter network was collected using seeds for known terrorist propaganda accounts in 2015. Model parameter adjustments were made to reproduce the collected network’s summary statistics, stylized facts and general structural measures. We produced an approximate network in line with the general properties of our collected data. We present our findings with a focus on the challenging aspects of this reproduction. We find that while it is possible to generate a social media network utilizing a few simple rules, numerous challenges arise requiring departure from the agent viewpoint and the development of more useful methods. We present numerous weaknesses and challenges in our reproduction and propose potential solutions for future efforts.

\textbf{Keywords:} social networks, social media, challenges, bottom-up growth, agent-based modeling, social simulation

\section{introduction}
\label{sec:intro}

Network science has grown in popularity in the last few decades, especially with the ubiquity of social media platforms and the Internet in general. However, much remains to be explored in relation to the processes by which networks form temporally, the underlying social "forces" that underpin their development, and thus the various distributions emerged by those social forces.

In this paper, we provide exploratory analysis and comparison of two networks—one is generated through an agent-based model utilizing a small number of simple rules, and a second network collected from the Twitter API using seed accounts for a terrorist organization used in an earlier paper to study propaganda dissemination ~\cite{Shaheen2015}.

This work considers three separate but equally important lines of inquiry–the first being that of computational and mathematical sociologists along the \cite{Travers1969}, \cite{Granovetter1973}, \cite{Freeman1977} path of research, where centrality measures were developed to provide an understanding of actors in networks--a microscopic viewpoint. The second line of inquiry is that of social physicists, epitomized by \cite{Price1976} and \cite{Barabasi1999} in which random networks were used to emerge network-level properties through implicit agent-level rule exploration. The third and final line of inquiry is that of the agent perspective \cite{Schelling1969,EpsteinAxtell1996} where agent-level behaviors or "rules" were used to grow and ultimately reproduce system-level properties and societal effects—a growing field.

In this paper, we consider a specific type of generative network model where new agents enter a network at a pre-determined rate proportional to the network's size at \textit{t-1} and where agents are limited by some constant $\kappa$, in their rate of tie formation for every time-period. We invoke simple agent-level rules known to be prevalent in the majority of social networks, and we attempt a statistically \textit{unprincipled} yet fruitful exploration of the collected network's properties. In particular, we discuss the challenges imposed on high-fidelity reproduction by limiting our efforts to agent only behaviors and argue that better reproduction must encompass both agent-level rules \textit{and} exogenous rules.

In this section, we will motivate the reader as to the importance of the challenges we present in this paper and draw from the literature important principles as an outline of later discussions. In \autoref{Methodology}, we discuss the model's internal workings and relevant agent rules and in later sections we elucidate the incident challenges of this model.

\subsection{Incumbent Motivation}
\label{Motivation}

Daesh, also known as ISIS (Islamic State of Iraq and Syria) is a profoundly impactful terror network with affiliate groups all over the world, and in particular the Middle East. The group has been responsible for dozens of terrorist attacks globally, as well as the subjugation of a local populace in Syria and Iraq. This paper will report efforts to reproduce a network with similar features and attributes to a previously collected one by \cite{Shaheen2015} drawn from known seed accounts belonging to the group. Reducing the group's reach and ability to coordinate and disseminate propaganda motivated this effort. Through precise reproduction one can identify the social dynamics by which groups develop on open social media platforms and can then prepare network-based interventions as a method of combat. Though we admit that the goal of reproducing social media networks is a general one and thus methods discussed in this paper can and often do apply to any social media network.

Agent-based simulation is a well-positioned collection of techniques, methods, and theories particularly amenable to the study of social networks and generally suitable where individuals or \textit{agents} govern the space of possible outcomes of the underlying system.

\subsection{Ideal Models}

Existing dynamic models of social networks can often be rigid in their assumptions and thus employ tractable, simple methods to describe how social networks form. For example, the well-known preferential attachment/cumulative advantage model \cite{Price1976,Barabasi1999} implicitly assumes that nodes, once entered a network, have an equal probability of connecting with a local node as they do with a node far away in the topology of the network so long as the probability of edge formation is linearly proportional to alter's degree centrality according to (\ref{eq:proprefmodel}). In agent-based modeling terms, the agents in the cumulative advantage model have no "vision" (or perfect vision), and thus are free to randomly choose any other agent to link to and consequently undergo a probability test to determine whether they will connect. 

In general, but more specifically for social media networks, while the former social process could be applicable it is rare, primarily because agents will tend to have better search opportunity in local topological space; whether said opportunity is emerged due to their own preferences or through a given platform's recommendations. This key feature (or drawback) is best represented through analysis by \cite{Klemm2002} in that while systems such as the Internet may in fact be better modeled by a cumulative advantage model, the model fails to reproduce the web's clustering coefficient properties, hence \textit{local} edge formation.

\begin{equation} \label{eq:proprefmodel}
p_{i} =  \dfrac{k_{i}}{\sum_{j=1}^{n}k_{j}} \text{\textit{, where n = number of nodes, k = degree}}
\end{equation} 

Furthermore, idealized models such as the cumulative advantage or the small world models \cite{Watts1998} fail to provide a realistic dynamic for edge formation, though perhaps that was never their objective. For example, one feature of an ever-growing cumulative advantage model is that average degree $\langle$k$\rangle$ scales with network size \textit{n}. While this may apply in principle, it simply cannot be the case that for every time unit \textit{t} and for every node \textit{i} the possible increase in node incremental out-degree is infinite. There is simply not enough time or an amenable social search mechanism that would allow ego to connect with an unlimited number of alters or vice versa. This is especially the case for backcloth networks under consideration in this paper and we consider it to be self-evident. 

\subsection{Statistical Models}

On the other hand, statistical models that ascertain extant social "forces", e.g. the body of work on exponential random graph models (ERGMs) \cite{Robins2007a} decompose a network into some collection of social processes by using structural counts of meso-scale motifs as the basis for statistical modeling of the network. Much interpretation is subject to the modeler’s assumptions and biases when utilizing these models as is the case with any other class of models including agent-based models. Thus, statistically principled models are still subject to limitations imposed by what has been previously proven to exists in a network in question. 

The strengths of statistical models are obvious--primarily following a theme of \textit{principality} in analysis. However, through their utilization assumptions are made which make the task of close and approximate reproduction of social media networks non-trivial. The first of which is that ERGM family optimizes for mainly the degree distribution of a given network, and assigns equal probability of emergence to all networks with that particular degree distribution. That is, the overall topological patterns of a given network are simply not emerged, with few notable exceptions. Equation (\ref{eq:ERGM}) defines the probability distribution of some realized graph \textit{g} on support $\varsigma$, conditioned on some parameter vector $\theta$, some arbitrary covariate set \textit{X}, representing a collection of social "forces", with some reference measure \textit{h(g,X)} usually taken to be equal to 1.

\begin{equation}
	\Pr( G = g|\theta, X) = \dfrac{exp(\theta^{T} t (g, X))}{\sum_{g' \in \varsigma} exp (\theta^{T} t(g',X) h(g',X)} h(g,X)
	\label{eq:ERGM}
\end{equation}
And, while for large \textit{n} the state of the art computational packages designed to deduce extant social forces struggle to compute parameters in any reasonable time-frame, the main critique we propose here is that topologically, the generated networks produced by simulations parameterized through deductive analysis rarely yield generated networks with similar topology to networks gathered from data; though they tend to approximate the intended degree distributions with superb precision.

\section{methodology}
\label{Methodology}
\subsection{Collected Network}
\label{Collected Nertwork}
A Twitter backcloth network (friend/follow) was collected using the Twitter REST API through the use of seed accounts to study terrorist propaganda in a previous study \cite{Shaheen2015}. Seed accounts were known to belong to terrorist propaganda operations of the Islamic State of Iraq and Syria (ISIS). The identified accounts were used to form the basis of a snowball sample. On average, seeds ranged in number from 15-20 accounts. The API collected up to 200 of their followers (in-degree) at first degree distance from ego, and subsequently those accounts were used as seeds to collect their first degree connections. The process was repeated for a total of 4 degrees of separation from seed nodes. Subsequently, analysis was conducted on the network with a focus on summary (network level) measures, features, and properties. \autoref{fig:wolf_network} shows a visualization of the collected network analyzed in this paper. 

\subsection{Simulated Network}\label{Simulated Network}
\subsubsection{Entry and Activity Rates} \label{sec:entryactivity}
The simulated model used simple rules to generate a directed network of roughly 160,000 nodes affording a 1:1 comparison with our collected network. The rule-set was under a time-constraint condition such that new nodes entering the network arrived at a much slower rate than the rate of edge formation of existing nodes and at a rate proportional to the current size of the network--a gravity model. We introduce this rule as \textit{time-based coupled rule-set}. That is, a set of rules which only act in unison and within proportional rates. The rule-set is mechanistically assigned to reflect the time constraints imbued within any action taken in a social media environment. 

Consider that each time unit of a simulation is a reflection of finite time in which agents may perform actions based on a given rule. It is easy to see that agents should be constrained by time. Furthermore, consider that in the absence of precise data, one must determine a rate of entry of new agents into the simulation. This rate affects which agents in a simulation will be more likely to receive ties from others and build ties with others, because they've had a greater opportunity to do so in the network (based on how long they've been active in the environment). In this case, a reasonable choice for rate of entry is one where we assume that the current size of the network at time \textit{t} determines how many new agents will enter the network at time \textit{t+1}. In other words, we conceive that as the network grows, more agents from outside the network learn of the existence the network and thus choose to enter the network. (\ref{eq:entryrate}) captures this dynamic.

Time should also constrain the number of actions each agent can take in the network environment. For each additional agent entering the network, and since agents in reality act in parallel, the total time-based actions available $\tau$ should also increase. In other words, there should be more general activity in the network given that there are more agents in the network overall. (\ref{eq:actiontime}) captures this dynamic. Consequently, it is trivial to show that (\ref{eq:actiontoentry}) describes the relation between both entry rule and action rule in terms of $\tau_{t+1}$,

\begin{equation} 
n_{t+1} = n_{t} (\nu+1) 
\label{eq:entryrate} 
\end{equation}

\begin{equation}
\tau_{t} = n_{t} \psi  
\label{eq:actiontime}
\end{equation} 

\begin{equation}
\tau_{t+1} = n_{t} \psi (\nu +1) 
\label{eq:actiontoentry}
\end{equation}

where $n_{t+1}$  is the number of nodes at \textit{t+1} and $\nu$ is rate of entry parameter, where $\nu$ $\in$ [0,1] and where $\tau_{t}$ is the number of actions available to be taken at each time interval for all nodes and $\psi$ is an action rate parameter, where $\psi$ $\in$ [1,$\infty$).

While time-constraining is justifiable based on basic intuition, we can further ground it on quantitative observations of friend/follow activity when compared with quantities of new accounts being created on Twitter, and the drastically differing rates by which different types of actions on social media platforms are taken (such as following, tweeting etc.) when compared with the rate of arrival of new nodes to the network. Primarily, the coupled rule-set signals our intent to mechanistically add new nodes to the simulation through a neutral method of addition that does not amount to undesirable social forces. The result of including this rule-set with underlying uniform probabilities of tie formation is a growing dynamic random graph with slight right-skew--a dynamic Erdos-Renyi devoid of non-random social forces. Thus, the rule-set serves the purpose of allowing us to model entry and activity in the network uncommittedly while focusing on creating non-random social forces that emerge the properties of data-based collected networks. Typical values of the entry parameter were an order of magnitude lower than the current size of the network, where typical values of the activity parameter were of an order of magnitude or more higher than network size.

\subsection{Agent Rules}
\label{Agent Rules}
If we are to consider every standard social force option provided by the exponential random graph modeling family we would be presented with tens of interesting possibilities for network reproduction. However, modeling, principled or otherwise, still relies on the modeler's judgment on choosing which social forces to include in a statistical analysis based on micro and meso-structures. Moreover, while interesting social forces might be discoverable through a deductive analysis of our data set, it is not a given that these forces, when combined, cannot be well-represented by a random behavior agent-based rule due to their small impact on structural emergence.

As indicated earlier, an important aspect of reproduction is the topological arrangement of the collected network. Therefore, agents are endowed with rules based on social forces which we know to exist in all but a few networks. Rules are activated in a sequence of activation as outlined below. In addition to the time constraints outlined earlier, agents are constrained on how many links they can build on each turn through a user-adjusted parameter. All probability tests for whether a new link is created are based on uniform distributions--a Bernoulli test. As outlined earlier we also subject all actions in the network to activity or time limitations, thus every probability test that determines whether a link will be created between any two nodes is modeled as an actual event. Socially, it is the equivalent of an agent seeing an alter's profile or tweet and then deciding to follow or not to follow them. Algorithm \ref{code:actioncalculation} summarizes this mechanism.

\begin{algorithm}[h]
		
	\For{turn t+1}{ numberofnewnodes(t + 1) = numberofnewnnodes(t) * (1 + $\nu$) \\
	numberofactions(t + 1) = numberofnewnodes(t + 1) * $\psi$ }
\caption{Calculate the number of nodes to be added and the total number of actions available.} 
\label{code:actioncalculation}
\end{algorithm}

\subsubsection{Rule: Randomness}
\label{Randomness}
Once an agent arrives (instantiated) into the network it undergoes a uniform probability test and based on the success or failure of the test a link is created with some randomly selected node. This rule is designed to increase the density of the network and to act as a placeholder for less impactful social forces not considered in our simulation. The rule is summarized in \autoref{code:randomness}.

\begin{algorithm}[h]
	\For{entrynode in newnodes AND actions != 0}{
	\eIf{p < bernoulli.random.test}
	{link entrynode to random node \\ 
	actions = actions - 2}
	{do nothing \\ actions = actions - 1}}
\caption{Agents entering the network will receive an initial opportunity to connect to any node already in the network with uniform probability.}
\label{code:randomness}
\end{algorithm}

\subsubsection{Rule: Triadic Closure}
\label{Triadic Closure}
The triadic closure rule ascertains whether a randomly selected agent's first degree connection has a mutual link with a third node and if so, according to a Bernoulli test, will then create a tie to the second node. This is executed in line with time constraints placed on network actions and amounts to connecting ego with a friend of a friend, if available (transitivity). It is important to note that transitivity is a social force that can occur through proxy social processes, including homophily (attribute), reciprocity (structural), and even classical preferential attachment. Therefore, by inclusion we use it as an aggregate descriptor of a number of other social forces, including classical transitivity (the friend of my friend becomes my friend). The mechanism is illustrated in algorithm \ref{code:transitivity}.

\begin{algorithm}[h]
	\For{randomnode in network AND actions != 0}{
		\eIf{p < bernoulli.random.test AND randomnode has neighbors AND randomneighbor has neighbor}
		{link randomnode to neighborofneighbor \\ 
			actions = actions - 3}{do nothing \\ actions = actions - 1}}
	\caption{Choose a random node that has neighbors then according to a Bernoulli test, link with a randomly selected neighbor of its own neighbors.}
	\label{code:transitivity}
\end{algorithm}

\subsubsection{Cumulative Growth}
\label{Cumulative Growth}
Cumulative growth, to be differentiated from preferential attachment, constructs a Bernoulli test to evaluate whether a random node connects to another random node if the second node has a higher in-degree. The rule is designed with a preferential attachment mindset but without explicitly defining the rule to behave in a manner equivalent to equation (\ref{eq:proprefmodel}). We assume that specifically on social media networks nodes do not always enter the network having any knowledge of the network's pre-existing structure or of the relative importance of a particular node. Agents only evaluate the connecting node based on their relative importance to ego not to the network as a whole. The rule/mechanism is further explained in algorithm \ref{code:cumulativegrowth}.

\begin{algorithm}[h]
	\For{firstrandomnode in network AND actions != 0}{
		\eIf{p < bernoulli.random.test AND secondrandomnode.indegree > firstrandomnode.indegree}
		{link firstrandomdnode to secondrandomnode \\ 
			actions = actions - 2}{do nothing \\ actions = actions - 1}}
		\caption{Select two random nodes. If the second node has a higher in-degree than the first node, conduct a Bernoulli test and if successful the first node can connect to the second.}
		\label{code:cumulativegrowth}
\end{algorithm}

\subsubsection{Distance-Assisted Closure}
\label{Distance-Assisted Closure}
The rule mechanism connects random nodes with high in-degree nodes if they are no more than a distance of 2 away and according to a Bernoulli test. Thus, the rule is a form of closure but it is more prevalent when an agent is within a shorter distance to a "powerful" agent. The rule represents the recommendation engine for Twitter users which seems to work by recommending popular accounts only within a small distance away, likely according to some community detection algorithm. The precise method by which link recommendations are made to users of the social media network is not made public.

\begin{algorithm}[h] 
	topnodes = list(search for top nodes by indegree)\\
	\For{randomnode in network AND actions != 0}{
		\eIf{p < bernoulli.random.test}
		{link randomnode to randomtopnode \\ 
			actions = actions - 2}{do nothing \\ actions = actions - 1}}
		\caption{Find the top nodes in the network. Select a random node. If a successful Bernoulli test is conducted, the random node links to a random top node.}
		\label{code:distanceclosure}
\end{algorithm}

\subsection{Model Development}
\label{Calibration}
All probability distributions used in the simulation were uniform in nature so as to not provide unnecessary complexity. To adjust the success rates of specific behaviors as they are applied to the nodal actions of those distributions, scalar parameters were used and adjusted based on qualitative evidence from Twitter's past reporting on the number of accounts and tweets on their platform \cite{Twitter2011}. 

\begin{figure} 	
	\begin{center}
		\begin{subfigure}[b]{0.3\textwidth}
			\centering
			\includegraphics[height=2.75cm]{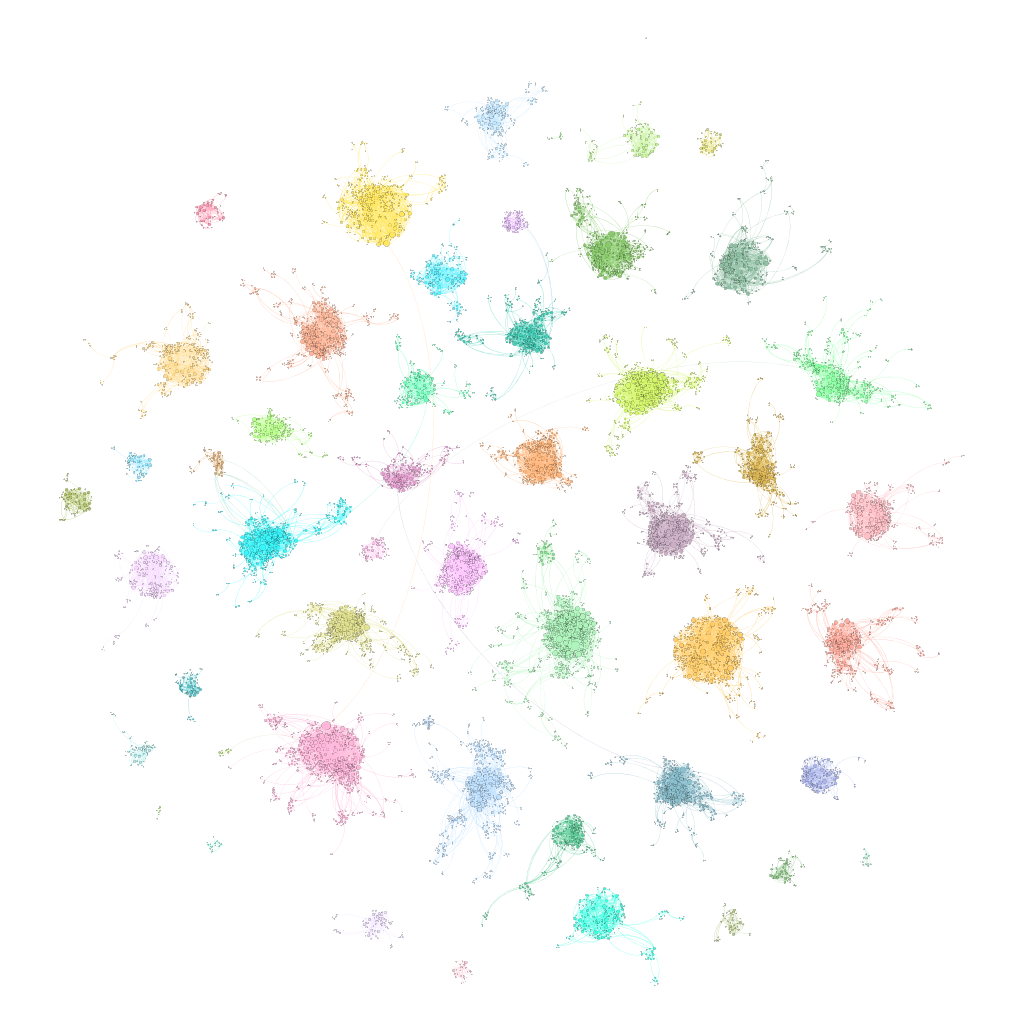}
			\caption{\centering N3: 5000 agents with random, cumulative and transitive behavior.}
			\label{fig:N3}
		\end{subfigure}
		\centering
		\begin{subfigure}[b]{0.3\textwidth}
			\centering
			\includegraphics[height=2.75cm]{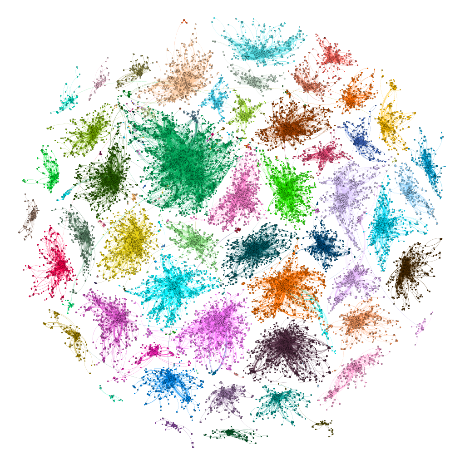}
			\caption{\centering N4: 10,000 agents with all but distance-assisted closure.}
			\label{fig:N4}
		\end{subfigure}
		\begin{subfigure}[b]{0.3\textwidth}
			\centering
			\includegraphics[height=2.75cm]{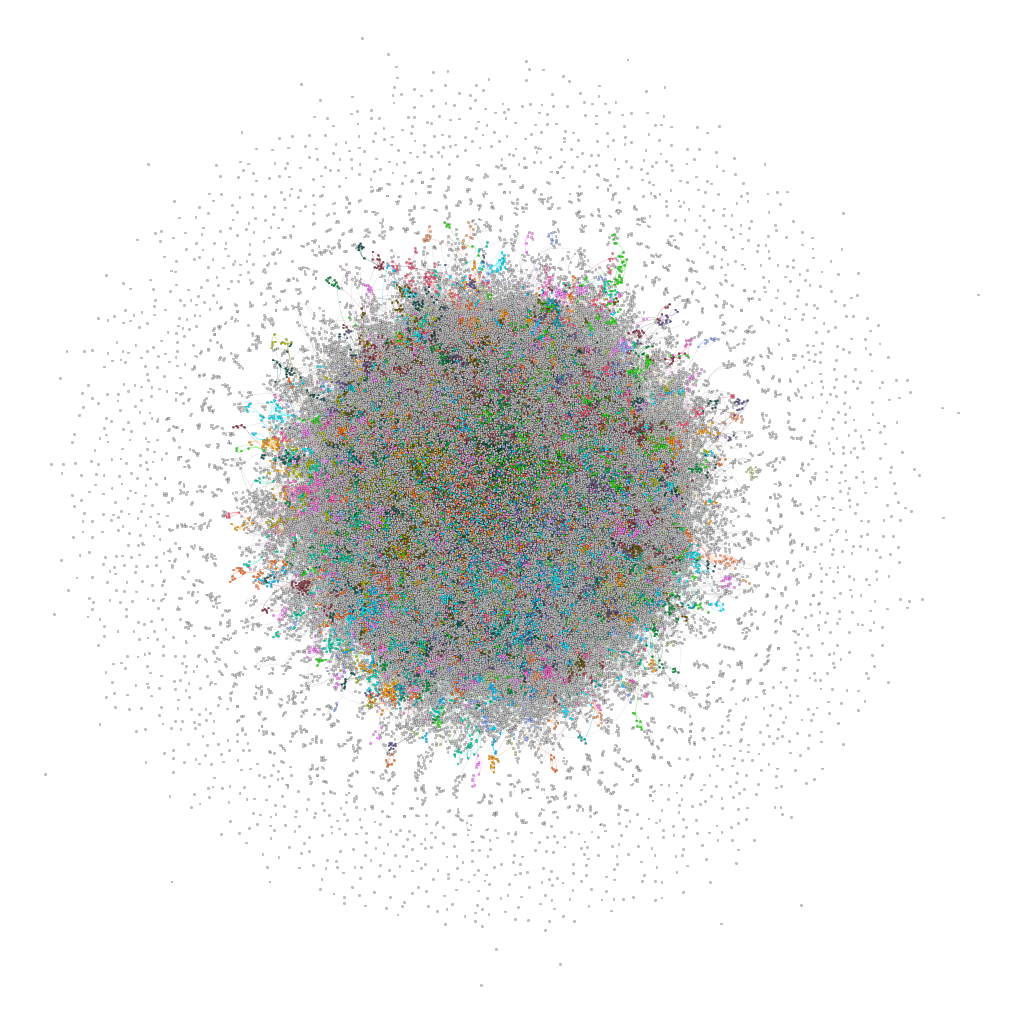}
			\caption{\centering N6: 80,000 agents with all rules active.}
			\label{fig:N6}
		\end{subfigure}
		\begin{subfigure}[b]{0.3\textwidth}
			\centering
			\includegraphics[height=2.75cm]{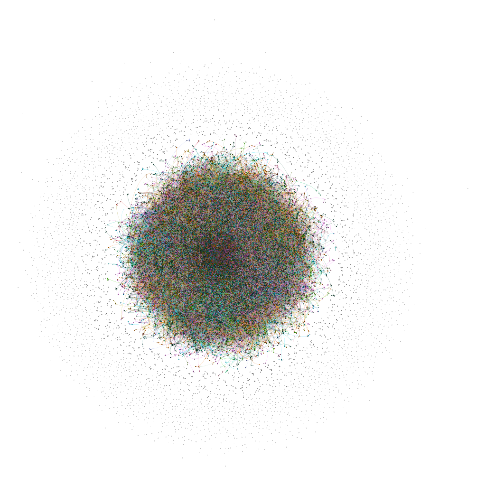}
			\caption{\centering N7: 160,000 agents with all rules active, without parameter adjustment.}
			\label{fig:N7}
		\end{subfigure}
		\begin{subfigure}[b]{0.3\textwidth}
			\centering
			\includegraphics[height=2.75cm]{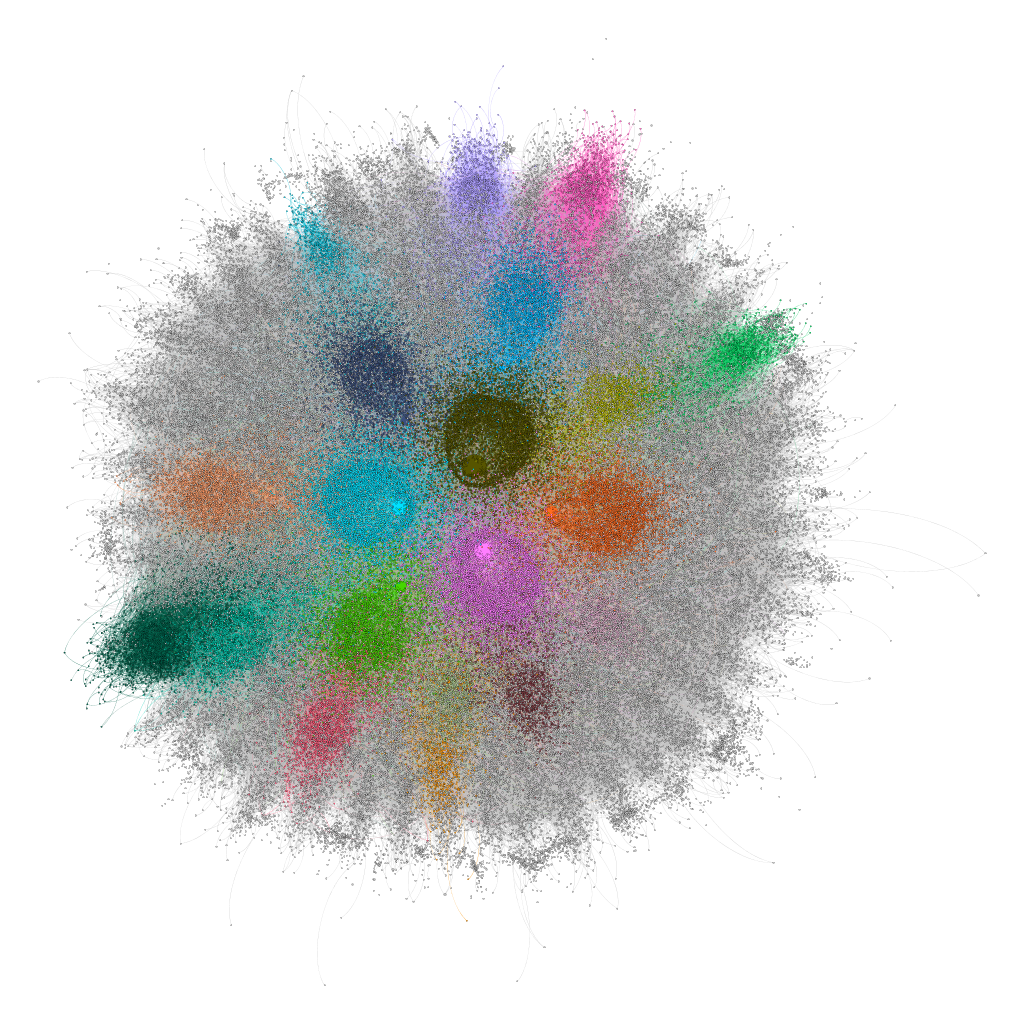}
			\caption{\centering N8: 160,000 agents. Final state of the network.}
			\label{fig:N8}
		\end{subfigure}
		\begin{subfigure}[b]{0.3\textwidth}
			\centering
			\includegraphics[height = 2.75cm]{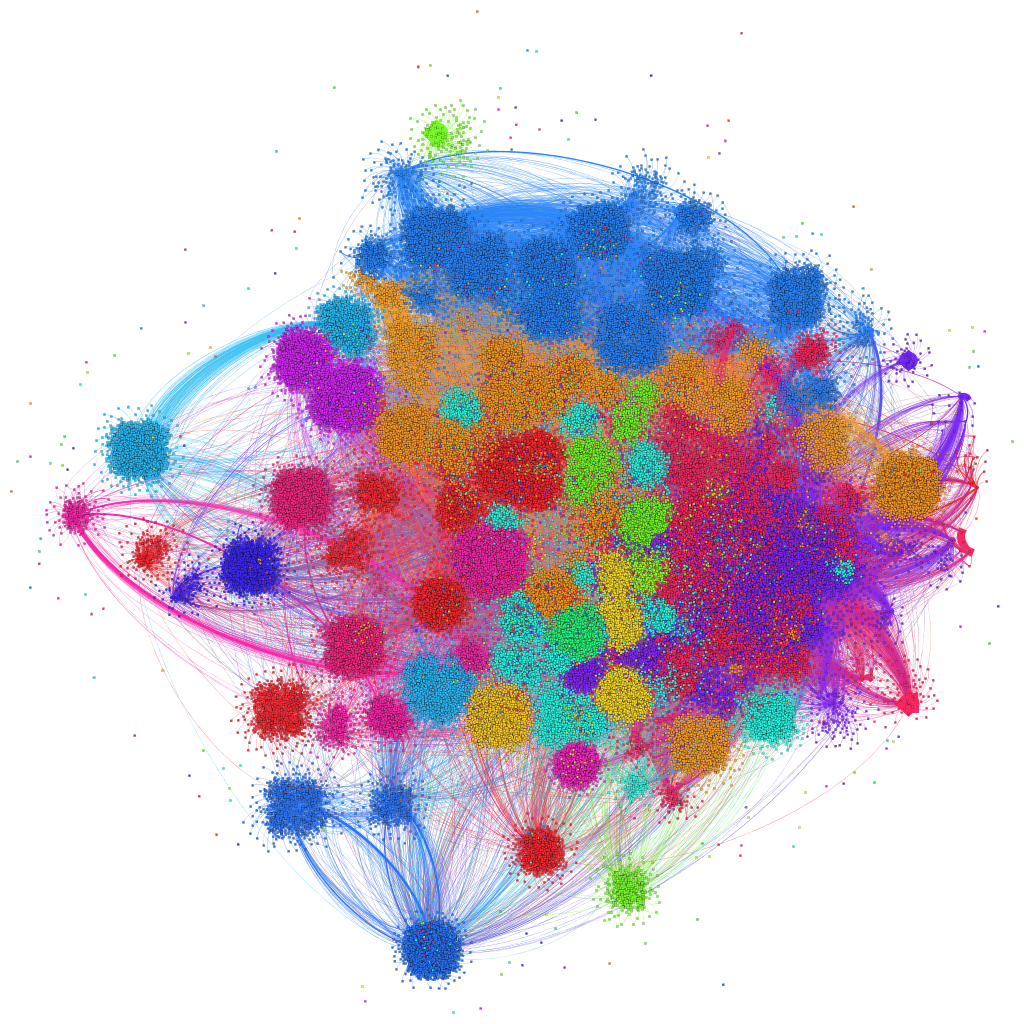}
			\caption{\centering The collected network used as a target for simulation.}
			\label{fig:wolf_network}
		\end{subfigure}
	\end{center}
	
	\caption{This figure showcases the various stages of model development for a typical run and their subsequent results N+number refers to synthetic networks in incremental developmental stages of node parameter adjustment.}
	\label{fig:modeldevelopment}
\end{figure}

For example, consider that the number of actions is analogous to the number of tweets. That is, an action does not always result in a follow, but in fact most tweets have little effect on the backcloth network in general. The Twitter report cited previously estimates that there were 50 Billion tweets with another report \cite{TSP2018} estimating that there were roughly 85-100 Million active users during a 2011 period. Using these values as a proxy for activity and network size, the difference between the level of activity on the network and the number of users must be an order of magnitude at minimum. These stylized facts serve as the underlying pillars of our choice of parameters. In order to incrementally develop the rules outlined in previous sections, an iterative process of development and testing was adopted. The model was repeatedly run for a network size of 50, 500, 1000, 5000, 10,000, 80,000, and finally 160,000 nodes. We continued to adjust the parameters in order to achieve a similar level of both total number of edges and total number of nodes. \autoref{fig:modeldevelopment} shows various stages of model development yielding a final state of the network.

\section{Results}
\label{Results}
\subsection{Imperfect Reproduction}

We find that this model was able to reproduce a similar network with corresponding properties, though with several exceptions related to path length and diameter of the network, which we found challenging to reproduce without investigating a larger portion of the possible parameter space. \autoref{tab:networkcomparison} compares summary measures of the collected network and the generated network.

\begin{table}[hb]
	\centering
	\caption{In this table we compare some of the summary measures of the collected network and the generated network.}
	\begin{tabular}{|l|l|l|}
		\hline
		Summary Statistic       & Collected & Generated \\ \hline
		Number of nodes         & 158,844   & 159,950   \\ \hline
		Number of Edges         & 504,441   & 397,198   \\ \hline
		Average Degree          & 6.351     & 4.976     \\ \hline
		Average Weighted Degree & 7.15      & 4.664     \\ \hline
		Network Diameter        & 10        & 21        \\ \hline
		Modularity              & 0.502     & 0.641     \\ \hline
		Average Path Length     & 119.5     & 80.12     \\ \hline		
	\end{tabular}
\label{tab:networkcomparison}
\end{table}

\begin{figure} 
\centering
\begin{subfigure}[b]{0.3\textwidth}
	\centering
	\includegraphics[height=4cm]{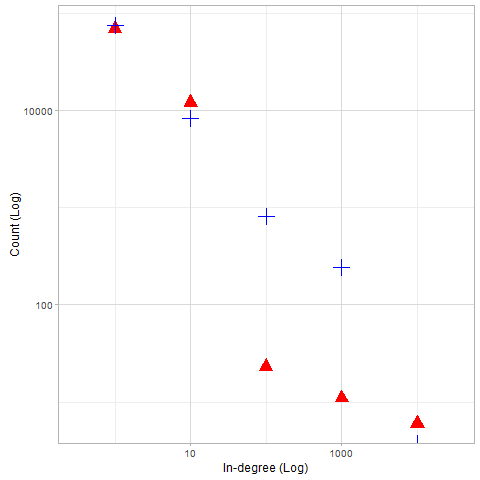}
	\caption{\centering In-degree (Log-Log)}
	\label{fig:indegreeloglog}
\end{subfigure} 
\begin{subfigure}[b]{0.3\textwidth}
	\centering
	\includegraphics[height=4cm]{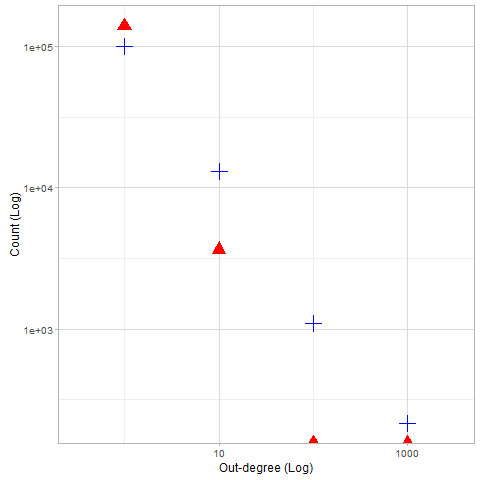}
	\caption{\centering Out-degree (Log-Log)}
	\label{fig:outdegreeloglog}
\end{subfigure}
\begin{subfigure}[b]{0.3\textwidth}
	\centering
	\includegraphics[height=4cm]{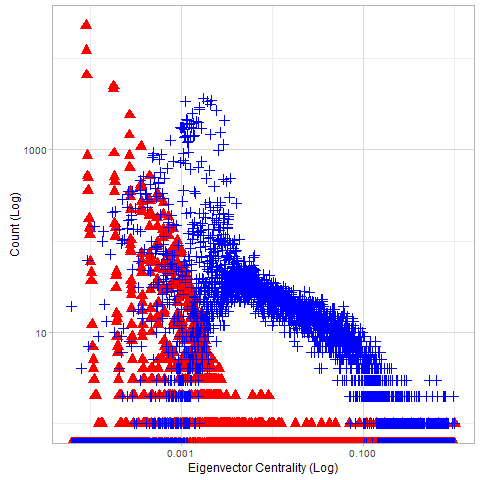}
	\caption{\centering Eigenvector (Log-Log)}
	\label{fig:evcloglog}
\end{subfigure}\\
\begin{subfigure}[b]{0.3\textwidth}
	\centering
	\includegraphics[height=4cm]{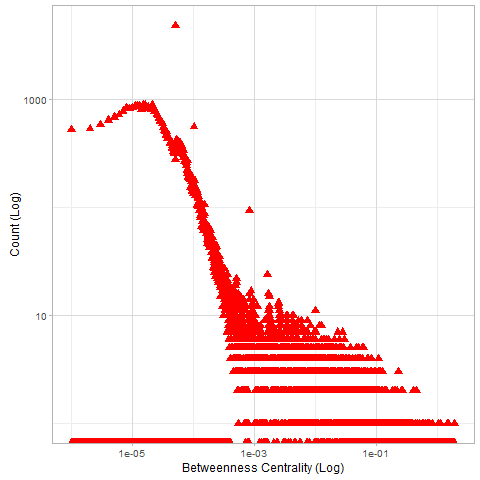}
	\caption{\centering Betweenness (Log-Log) }
	\label{fig:betloglog}
\end{subfigure}
\begin{subfigure}[b]{0.3\textwidth}
	\centering
	\includegraphics[height=4cm]{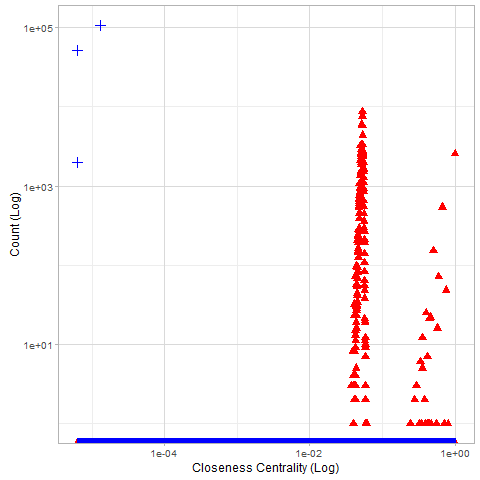}
	\caption{\centering Closeness (Log-Log)}
	\label{fig:clologlog}
\end{subfigure}\\

\caption{Log-log plots are shown for various centrality distributions. Blue(+) data points represent the collected network (data). Red ($\bigtriangleup$) represents the generated (simulated) network. Null values on log-log were removed.}
\label{fig:modeldistributions}

\end{figure}

We also find that we were able to reproduce the general shape and outline of both the in-degree (\autoref{fig:indegreeloglog}) and out-degree (\autoref{fig:outdegreeloglog}) distributions, and that many of the summary statistics are in close proximity to intended values with some exceptions.  The remainder of \autoref{fig:modeldistributions} compares the distribution of the network produced by our model versus data. Visually, similar properties of the various measure distributions were confirmed for both the collected network and generated network when comparing \autoref{fig:N8} with \autoref{fig:wolf_network}.

\subsection{Distributions' Properties}
\label{sec:distprop}
The model performed well in certain areas. For example, in the lower values of the in-degree distribution the model approximates expectation closely (\autoref{fig:indegreeloglog}). However, in the mid-section of in-degree values we report digressions. This is perhaps due to the generated network's lower density when compared with the collected network. This same pattern applies to the out-degree distribution (\autoref{fig:outdegreeloglog}) where higher and lower values are closely reproduced but divergence in the mid-values appears. Overall, the model under-performs in expected mean degree. Additional exploration of the model's parameters will likely yield a better result in this arena.

Divergence in the mid values of the degree distribution is confirmed by the distribution of the eigenvector centrality (\autoref{fig:evcloglog}), where the model's output contains an almost supralinear rate of decay with fewer nodes having mid-level or high-level values than data. Moreover, the complexity of the data's eigenvector distribution provides key evidence pointing to possible  exclusion of certain social forces--and consequently rules--from the model. Likely, these unincorporated rules have strong influence on the behavior of nodes when they reach a certain age--or in structural terms--a certain mean degree. 

The difference in the generated network's and the collected network's diameter likely gives the best hint for missing rules (\autoref{tab:networkcomparison}). Networks that maintain clustering while possessing shorter distances (path lengths) are generally known to be governed by small world behavior. That is--nodes in these networks find more optimal pathways by re-wiring their existing links. Our model contains no re-wiring rules. Once a node gains ties those ties are never reconfigured. 

Furthermore, it is possible that a re-wiring rule-set may not take on a classical form, but that it could be a result of another social force. For example, our model does not account for tie decay or removal nor does it model agents leaving the environment (thereby removing all of ego's outgoing and incoming ties). Social processes like the former would likely have a small world effect and shorten overall distances. It would be ever more interesting if an omitted small world rule was applicable to the mid-range of the degree distribution, describing some tie maintenance mechanism or mean degree local maximum to reflect Twitter's policy of allowing users to only follow a certain number of users. 

Curiously, another platform-based policy might explain the differing betweenness (\autoref{fig:betloglog}) and closeness (\autoref{fig:clologlog}) centrality distributions. As observed, albeit a few nodes with high values, almost all betweenness and closeness values for our collected data are close to zero. This is likely due to the way Twitter reports account output through the API by limiting the reporting to only 200 accounts for each seed. In other words, Twitter's API prunes intermediary links in the data set by restricting reporting to only a fraction of the network's topology. 

\section{Challenges and Opportunities}
\label{Discussion}
What the results show is an informative exercise in bottom-up growth of social networks. With additional tuning, extensions, and modification to the structure of the agent-based model, including the additions of a rule to govern small-world behavior, better results can be attained. As results show, with only a few simple rules, we were able to reproduce the general properties of a social media network, though much remains to be explored in the reproduction of very specific patterns. A number of challenges emerge from this exercise that can be generalized to the modeling of social media networks through an agent viewpoint. We summarize those challenges as follows.

\subsection{Entry and Exit Dynamics}

While many models do not account for entry and exit dynamics and focus on the modeling of edge formation, we find that entry and exit play an important role in the formation of a social media network, especially in our mode of modeling which explicitly represents time as a rule-set that constrains behavior. However, current literature leaves much to be desired in identifying possible schemes by which entry and exit manifests in dynamic networks. In our model, we simply state an assumption which seems reasonable from a sociological point of view and one which would ensure that the network formed through only the entry rule is devoid of social forces, which we then added manually and based on our expectations. Moreover, entry and exit dynamics are important for any agent-based simulation where time and age of agents has some effect on behavior. In our model we coupled the rate of activity and entry to network size, but one can plausibly find other entry regimes that can produce a similar result. The reader should also note that entry and exit processes likely exhibit different properties from well-studied activation regimes--a collection of behavioral rules that determine which agent "goes first"--and not considered as relevant to our model at this time. 

\subsection{Departure From the Agent Viewpoint}

Almost immediately it was apparent that restricting our model to a structural viewpoint of networks (agents did not possess attributes) required departure from agent-only rules. This is most notable when low clustering of early iterations of the model necessitated the addition of a transitivity rule. Transitivity occurs in networks through other social forces (homophily, reciprocity etc.) as well as through an explicit "friend of my friend is my friend" process. But transitivity by itself is not a strictly agent-based rule--it requires that multiple sequential sub-processes occur (A connects with B, B connects with C, B introduces A to C, A connects with C) or that agents are able to learn about the network's topology beyond their immediate first degree connections--the addition of "vision"--ensuring closure of open triads.

Additionally, platforms' recommendation engines play a larger role in how the network forms, by suggesting to users who they should connect with, what they should discuss, and what information to receive. The agent is not entirely driven by personal choices. In an applied sense, recommendations allow some node A to connect with some node B based on no discoverable agent-rule which we can simulate. For example, consider a hypothetical example where a social media platform discovered that there is some positive relation between the number of n-cycles (perhaps most notable at n = 9) and advertisement click-through rates. Consequently, they develop a recommendation engine that analyzes the entire network and then makes link suggestions to users in order to increase the number of 9-cycles on their platform. How would this be modeled without departure from a strict agent perspective? After all, to produce a specific 9-cycle, agent linkages must become deterministic--each agent needs to know which specific agent to connect to next. Thus, it is not immediately clear that a strict agent rule-set can reproduce the full properties of any social media network that employs meso-scale recommendations such as transitive link building. 

\subsection{Modeling Time and Search}

In a typical attempt at reproduction through simulation, time is often modeled as an abstract exogenous variable--sometimes providing an aesthetic appeal without actually integrating useful dynamics. In idealized, mathematically soluble models--time is often not considered entirely. Yet, social behavior is subject to time constraints and must have some linkage to levels of activity, as we have tried to show in this model. Therefore, we guesstimated that linkage by evaluating the proportion of all activity on Twitter with respect to the entry of nodes and formation of links for the backcloth (friend and follow) network. In fact, the model represents a natural analogy to account creation, tweeting, and following others if we consider every failed Bernoulli test equivalent to an agent observing another agent's tweet and deciding not to follow said agent. However, the method we use to estimate this mechanism is rudimentary at best, and without very detailed usage data only available via Twitter's internal dataset or statistical estimation procedures gathered from samples the task becomes non-trivial.

Social search represents another related challenge, though social search mechanisms have generally been well-investigated in the literature. Albeit, social search mechanisms are varied and are highly subject to a platform's digital culture. Moreover, search mechanism in the literature are rarely subject to time constraints, making calibration efforts that are of great importance to this class of models challenging. 

\subsection{Comparisons: Knowing When we Get it Right}

Finally, we find the process of validating results gained from network simulations to be insufficient for reaching any principled conclusion, especially in evaluating the topological arrangement of either data or simulation. Non-parametric statistical tests of the produced distributions or the use of hamming distance as a comparator largely do not capture the extent for which a network has been reproduced, especially in structural terms. To conduct a comparison for this paper we relied on summary measures and a qualitative, visual comparisons, but even if we utilized classical statistical techniques, this class of simulations generally run in exponential time making large numbers of runs--necessary for good sampling techniques--a practical impossibility. Furthermore, sample-based methods of validation would still not suffice in evaluating the topology of networks produced since no principled method exists for "averaging" topology.

\section{Conclusion}
\label{Conclusion}

In this paper we showed that a generated network can produce summary statistics and distributions comparable to that of a collected network from Twitter, though with many unsolved challenges and opportunities for further investigation and ultimately, personal limitations of the author's knowledge. We showed that the use of a coupled-time rule-set is complementary to capturing the precise rates by which a social media network can be reproduced. We showed that a small number of simple rules, based on nothing more than Bernoulli tests can be integrated into an agent-based model to reproduce a highly complex social media network with multiple local forces acting.

Additionally, through identifying the most salient challenges and critiques of our reproduction we implicitly propose a path forward to improving the model and reaching a more precise reproduction of social media networks--a first step in exploring dynamic structural interventions to prevent online extremism.

\bibliography{smalllibrary}
\bibliographystyle{IEEEtran}

\section*{Author Biography}

\textbf{\uppercase{Joseph A.E. Shaheen}} is a PhD candidate in Computational Social Science at George Mason University's Department of Computational and Data Science. 
Author email: jshaheen@gmu.edu

\end{document}